\documentclass[a4paper,12pt]{article}
\newtheorem{definition}{{\bf Definition}}[section]
\newtheorem{theorem}[definition]{{\bf Theorem}}
\newtheorem{lemma}[definition]{{\bf Lemma}}
\newtheorem{proposition}[definition]{{\bf Proposition}}
\newtheorem{corollary}[definition]{{\bf Corollary}}

\newtheorem{assumption}{{\bf Assumption}}

\usepackage{color}
\usepackage{ulem}

\def\<{\langle}
\def\>{\rangle}
\def\proof{\noindent{Proof. }}

\def\endproof{\hfill $\square$ \vspace{10pt}}
\def\H{{\mathcal H}}

\def\im{{\rm i}}
\def\e{{\bf e}}
\usepackage{amscd,amsmath,amssymb,graphicx}

\begin{document}

\title{Unitary equivalence classes of split-step quantum walks}
\author{Akihiro Narimatsu, \quad
Hiromichi Ohno, \quad 
Kazuyuki Wada}

\date{}

\maketitle

\begin{abstract}
This study investigates the unitary equivalence of split-step quantum walks (SSQW).
We consider a new class of quantum walks which includes all SSQWs.
We show the explicit form of quantum walks in this class, and
clarify their unitary equivalence classes.
Unitary equivalence classes of Suzuki's SSQW are also given. \\

\end{abstract}

\section{Introduction}

Quantum walks are analogous to classical random walks. 
They have been studied in various fields, 
such as quantum information theory and quantum probability theory. 
Especially, quantum walks give exponential algorithmic speed-ups.
There are various examples of quantum speed-ups via quantum walks. 
Comprehensive reviews of quantum walks are summarized in \cite{P, V}.

A quantum walk is defined by a pair $(U, \{H_x\}_{x\in V})$, in
which $V$ is a countable set, $\{H_x\}_{x\in V}$ is a family of separable Hilbert spaces,
and $U$ is a unitary operator on $\H = \bigoplus_{x\in V} \H_x$ \cite{SS2}.
In this paper, we discuss
split-step quantum walks (SSQW), in which $V ={\mathbb Z}$ and $\H_x = {\mathbb C}^2$ 
for $x \in {\mathbb Z}$.
These have been the subject of some previous studies 
\cite{CGGSVWW, FFS, FFS2, FFS3, M, S2, ST, T}.

There exist two definitions of SSQW.
One is defined by Kitagawa \cite{Ki1, Ki2},
and the other is defined by Suzuki \cite{FFS}.
As seen in (\ref{eq:kitagawa}) or \cite{FFS2}, Suzuki's definition is a generalization of Kitagawa's one.
However, we can still consider several generalizations of these SSQW.
Thus, we want to handle a class of quantum walks which includes all such generalizations.
To do this, we treat a unitary $U$ on $\H$ such that
$U\H_x \subset \H_{x-1} \oplus \H_x \oplus \H_{x-1}$ and 
${\rm rank}P_{x\pm 1} UP_x \le 1$, where $P_x$ is a projection onto $\H_x$.
Kitagawa's and Suzuki's SSQW satisfy these conditions.
On the other hand, an explicit form of $U$ is not clear. 
Hence, we show it in Sect. \ref{sect2}.

It is important to clarify when two quantum walks are 
unitarily equivalent in the sense of \cite{O, SS2}. 
If two quantum walks are unitarily equivalent, 
many properties of their quantum walks are the same. 
For example, digraphs, dimensions of Hilbert
spaces, spectrums of unitary operators, probability distributions of quantum
walks, etc. would be the same for each quantum walk. 
One of the aim of this paper is to determine the unitary equivalence classes of SSQW.
Then, we only need to study representatives of unitary equivalence classes
to know the above properties.
This helps studies of SSQW, 
because we can reduce the parameters 
which are needed to define quantum walks \cite{GKD, IKB, O4, O, O2, O3}.

In Sect. \ref{sect3}, we calculate unitary equivalence classes of 
the above quantum walks.
Moreover, we show unitary equivalence classes of Suzuki's SSQW in Sect. \ref{sect4}.
Chiral symmetry is also considered.

\section{Structure of SSQW}\label{sect2}

For each $x \in {\mathbb Z}$, $\H_x$ is a two dimensional Hilbert space ${\mathbb C}^2$,
and $P_x$ is a projection from $\H=  \bigoplus_{y\in {\mathbb Z}} \H_y$ onto $\H_x$.
$\{\e_1^x, \e_2^x\}$ is a canonical basis of $\H_x$.
$\H$ is also written as $\H = \ell^2 ({\mathbb Z}) \oplus \ell^2({\mathbb Z})$.
Strictly speaking, $\bigoplus_{x\in {\mathbb Z}} \H_x$ and 
$\ell^2 ({\mathbb Z}) \oplus \ell^2({\mathbb Z})$ are unitarily equivalent,
but we identify them.
The left-shift operator on $\ell^2({\mathbb Z})$ is denoted by $L$.

A quantum walk on $\H$ is a unitary operator, 
and there are several notations of a quantum walk $U$.
Here, we introduce three notations.
One is expressed as follows. 
Let $S$ be a unitary operator on 
$\H = \ell^2 ({\mathbb Z}) \oplus \ell^2({\mathbb Z})$ defined by
\[
(S\Psi)(x) = 
\begin{bmatrix} p_x \Psi_1(x) + q_{x-1} \Psi_2 (x+1) \\ 
\bar{q}_{x} \Psi_1(x-1) - \bar{p}_{x-1} \Psi_2(x) \end{bmatrix}
\]
for $x \in {\mathbb Z}$, $p_x, q_x \in {\mathbb C}$ and 
a vector $\Psi = \begin{bmatrix} \Psi_1 \\ \Psi_2 \end{bmatrix}  \in \H$.
Let 
\[
C(x) = \begin{bmatrix} \bar{a}_x & \bar{b}_x \\ b_x & - {a}_x \end{bmatrix}
\]
for $x \in {\mathbb Z}$ and some $a_x, b_x \in {\mathbb C}$,
and let $C$ be a unitary operator on $\H$ gived by
\[
(C\Psi) (x) = C(x) \Psi(x)
\]
for all $x \in {\mathbb Z}$. Then, we define a unitary operator $U$ by
\[
U = SC.
\]
Another notation is as follows. Let $p = (p_x)_{x \in {\mathbb Z}}$ and 
$q = (q_x)_{x \in {\mathbb Z}}$ be multiplication operators on $\ell^2({\mathbb Z})$,
and let
\[
S = \begin{bmatrix} p & qL \\ L^*q^* & -L^* p^* L \end{bmatrix}.
\]
$C$ is the same as above. Then, we define a unitary operator $U$ by
\begin{equation}\label{second}
U = SC =  \begin{bmatrix} p & qL \\ L^*q^* & -L^* p^* L \end{bmatrix}
\begin{bmatrix} a^* & b^* \\ b & - a \end{bmatrix},
\end{equation}
where $a = (a_x)_{x\in {\mathbb Z}}$ and $b = (b_x)_{x\in {\mathbb Z}}$ are
multiplication operators on $\ell^2({\mathbb Z})$.
The other uses the Dirac notation. We define $U$ by
\[
U = \sum_{x\in {\mathbb Z}} | 
p_x \e_1^x + \bar{q}_{x} \e_2^{x+1} \> \<{a}_x \e_1^x + {b}_x \e_2^x|
+ |q_{x-1} \e_1^{x-1} - \bar{p}_{x-1} \e_2^x\>\< \bar{b}_x \e_1^x - \bar{a}_x \e_2^x|.
\]
By straightforward calculation, we can see that the above three $U$'s are the same. 
The third notation is useful when we consider unitary equivalence classes of quantum walks. 
In the following, we use the third notation, mainly.

The definition of split-step quantum walk (SSQW) by Kitagawa 
(replacing $\theta_1/2$ and $\theta_2/2$ with $\theta_1$ and $\theta_2$)
is
\begin{equation}\label{eq:kitagawa}
U = 
\begin{bmatrix} I & 0 \\ 0 & L \end{bmatrix}
\begin{bmatrix} \cos \theta_1 & -\sin \theta_1 \\ 
\sin \theta_1 & \cos \theta_1 \end{bmatrix}
\begin{bmatrix} L^* & 0 \\ 0 & I \end{bmatrix}
\begin{bmatrix} \cos \theta_2 & -\sin \theta_2 \\ 
\sin \theta_2 & \cos \theta_2 \end{bmatrix} 
\end{equation}
on $\H = \ell^2({\mathbb Z}) \oplus \ell^2({\mathbb Z})$ 
for some $\theta_1, \theta_2 \in {\mathbb R}$.
This is unitarily equivalent to 
\begin{align*}
\begin{bmatrix} 0 & 1 \\ 1 & 0 \end{bmatrix}  U
\begin{bmatrix} 0 & 1 \\ 1 & 0 \end{bmatrix} 
=
\begin{bmatrix} \sin\theta_1 & \cos\theta_1L \\
\cos \theta_1L^* & - \sin\theta_1 \end{bmatrix}
\begin{bmatrix} -\sin\theta_2  & \cos \theta_2 \\
\cos\theta_2 & \sin\theta_2
\end{bmatrix}.
\end{align*}
This is expressed as \eqref{second}.
Suzuki generalized this, and defined SSQW by
\[
U = SC =  \begin{bmatrix} p & qL \\ L^*q^* & -L^*pL \end{bmatrix}
\begin{bmatrix} a & b^* \\ b & - a \end{bmatrix}
\]
for $p_x, a_x \in {\mathbb R}$, $q_x, b_x \in {\mathbb C}$ 
with $p_x^2 + |q_x|^2 =1$ and $a_x^2 + |b_x|^2 =1$.
Remark that this formulation is useful to consider chiral symmetry.

There can be several generalizations of these definitions.
Thus, we want to handle a class of quantum walks which includes all such generalizations.
For this purpose,
we treat a unitary $U$ on $\H$ which satisfies 
\begin{equation}\label{eq:rangecond}
U\H_x \subset \H_{x-1} \oplus \H_x \oplus \H_{x+1}
\end{equation}
and 
\begin{equation}\label{eq:rankcond}
{\rm rank} P_{x\pm 1} U P_x \le 1
\end{equation}
for all $x \in {\mathbb Z}$.
It is easy to see that Kitagawa's and Suzuki's SSQW satisfy these conditions.
On the other hand, the explicit form of $U$ like \eqref{second} is not clear.
Therefore, we will clarify it in the following.

\begin{lemma}\label{lem:eta}
For each $x \in {\mathbb Z}$, 
there exists an orthonormal basis $\{\eta_1^x, \eta_2^x\}$ such that
\[
U \H_x \subset {\mathbb C} \eta_1^{x-1} \oplus \H_x \oplus {\mathbb C} \eta_2^{x+1}.
\]
\end{lemma}
\proof
When ${\rm rank} P_{x} U P_{x+1} = 1$, 
there exist a unit vector $\eta_1^{x} \in \H_{x}$
such that 
\[
{\rm Ran} P_{x} U P_{x+1} = {\mathbb C} \eta_1^{x}.
\]
Similarly,
when ${\rm rank} P_{x} U P_{x-1} = 1$,
there exists a unit vector $\eta_2^{x} \in \H_{x}$
such that 
\[
{\rm Ran} P_{x} U P_{x-1} = {\mathbb C} \eta_2^{x}.
\]
Since $U$ is a unitary operator, $U\H_{x+1}$ and $U\H_{x-1}$ are orthogonal.
Therefore, if ${\rm rank} P_{x} U P_{x\pm 1} = 1$, 
we can prove that $\eta_1^x$ and $\eta_2^x$ are orthogonal by using \eqref{eq:rangecond}.
Hence, $\{\eta_1^x, \eta_2^x\}$ is an orthonormal basis of $\H_x$.
When ${\rm rank} P_{x} U P_{x+1} = 0$ or ${\rm rank} P_{x} U P_{x-1}=0$,
we choose $\eta_1^{x}$ and $\eta_2^{x}$ such that 
$\{\eta_1^x, \eta_2^x\}$ is an orthonormal basis of $\H_x$.
Then, we have the assertion.
\endproof

\begin{proposition}\label{prop:exclude}
If $UP_y$ is represented as
\[
UP_y= |\eta_1^{y-1}\>\< \zeta_1^y| + |\eta_2^{y}\>\<\zeta_2^y|
\]
for some $y \in {\mathbb Z}$ and orthonormal basis $\{\zeta_1^y, \zeta_2^y\}$ of $\H_y$,
then there exists an orthonormal basis $\{\zeta_1^x, \zeta_2^x\}_{x\in{\mathbb Z}}$ of $\H$
with $\zeta_1^x, \zeta_2^x \in \H_x$
such that $U$ is written as
\begin{equation}\label{eq:exclude}
U = \sum_{x \in{\mathbb Z}}  |\eta_1^{x-1}\>\< \zeta_1^x| + |\eta_2^{x}\>\<\zeta_2^x|.
\end{equation}
Similarly, if  $UP_y$ is represented as
\[
UP_y= |\eta_1^{y}\>\< \zeta_1^y| + |\eta_2^{y+1}\>\<\zeta_2^y|
\]
for some $y \in {\mathbb Z}$ and orthonormal basis $\{\zeta_1^y, \zeta_2^y\}$ of $\H_y$,
then there exists an orthonormal basis $\{\zeta_1^x, \zeta_2^x\}_{x\in{\mathbb Z}}$ of $\H$
with $\zeta_1^x, \zeta_2^x \in \H_x$
such that $U$ is written as
\begin{equation}\label{eq:exclude2}
U = \sum_{x \in{\mathbb Z}}   |\eta_1^{x}\>\< \zeta_1^x| + |\eta_2^{x+1}\>\<\zeta_2^x|.
\end{equation}
\end{proposition}
\proof
Assume that $UP_y$ is represented as
\[
UP_y= |\eta_1^{y-1}\>\< \zeta_1^y| + |\eta_2^{y}\>\<\zeta_2^y|.
\]
Then, $U\H_y = {\rm Ran}UP_y = {\mathbb C} \eta_1^{y-1} \oplus {\mathbb C} \eta_2^y$.
Since $U\H_{y-1} \subset {\mathbb C} \eta_1^{y-2} \oplus \H_{y-1} \oplus {\mathbb C} \eta_2^y$
and $U\H_{y-1} \perp U\H_y$,
\[
U\H_{y-1} = {\mathbb C} \eta_1^{y-2}  \oplus {\mathbb C} \eta_2^{y-1}.
\]
Therefore, there exists an orthonormal basis $\{\zeta_1^{y-1}, \zeta_2^{y-1}\}$ 
of $\H_{y-1}$ such that
\[
UP_{y-1} = |\eta_1^{y-2}\>\< \zeta_1^{y-1}| + |\eta_2^{y-1}\>\<\zeta_2^{y-1}|.
\]

On the other hand, 
\[
U\H_{y+1}^\perp = 
U\left( \bigoplus_{x \neq y+1} \H_x  \right)  \subset
{\mathbb C} \eta_2^y \oplus {\mathbb C}\eta_1^{y+1} \oplus \bigoplus_{x \neq y, y+1} \H_x.
\]
Hence, 
\[
U\H_{y+1} \supset {\mathbb C} \eta_1^y \oplus {\mathbb C} \eta_2^{y+1}.
\]
Since ${\rm dim} \H_{y+1} =2$, the above inclusion is an equation.
Therefore, there exists an orthonormal basis $\{\zeta_1^{y+1}, \zeta_2^{y+1}\}$ such that
\[
UP_{y+1} = |\eta_1^{y}\>\< \zeta_1^{y+1}| + |\eta_2^{y+1}\>\<\zeta_2^{y+1}|.
\]
Then, we can prove the assertion inductively.

The remaining part can be proven similarly.
\endproof

When $U$ is expressed as \eqref{eq:exclude}, $U$ is unitarily equivalent to
\[
U'= \begin{bmatrix} I & 0 \\ 0 & L \end{bmatrix}
\begin{bmatrix} a^* & b^* \\ 
b & -a \end{bmatrix}.
\]
This is similar to the first half of the Kitagawa's definition \eqref{eq:kitagawa}.
Similarly, \eqref{eq:exclude2} is unitarily equivalent to 
a modification of the second half of \eqref{eq:kitagawa}.
In this paper, we consider split-step quantum walks,
but $U$ with the form \eqref{eq:exclude} or \eqref{eq:exclude2} is not ``split-step''.
Therefore, we will exclude the cases \eqref{eq:exclude} and \eqref{eq:exclude2}
in the following.
Concretely, we will use the next assumption.

\begin{assumption}
There does not exist $y \in {\mathbb Z}$ such that
$UP_y$ is represented as
\[
UP_y= |\eta_1^{y-1}\>\< \zeta_1^y| + |\eta_2^{y}\>\<\zeta_2^y|
\quad {\rm or} \quad 
UP_y= |\eta_1^{y}\>\< \zeta_1^y| + |\eta_2^{y+1}\>\<\zeta_2^y|
\]
for some orthonormal basis $\{ \zeta_1^y, \zeta_2^y\}$ of $\H_y$.
\end{assumption}

\begin{lemma}\label{lem:rank0}
If 
${\rm rank} P_x U P_{x+1} =0$, then there exists a unit vector $\zeta_1^x \in \H_x$
such that 
\[
U \zeta_1^x = \eta_1^x.
\]
Similarly, 
If ${\rm rank} P_x U P_{x-1} =0$, then there exists a unit vector $\zeta_2^x \in \H_x$
such that $U \zeta_2^x = \eta_2^x$.
\end{lemma}
\proof
When ${\rm rank} P_x U P_{x+1} =0$, by \eqref{eq:rangecond} and Lemma \ref{lem:eta}, we have
\[
U \H_x^\perp = U \left( \bigoplus_{y\neq x} \H_y \right) \subset 
{\mathbb C} \eta_2^x \oplus \bigoplus_{y\neq x} \H_y .
\]
This implies 
\[
U\H_x \supset \left({\mathbb C} \eta_2^x \oplus \bigoplus_{y\neq x} \H_y \right)^\perp
=  {\mathbb C} \eta_1^x.
\]
Hence, there exists a unit vector $\zeta_1^x \in \H_x$ such that $U \zeta_1^x = \eta_1^x$.

We can prove the remaining part similarly.
\endproof

\begin{proposition}
For each $x \in {\mathbb Z}$,
there exists an orthonormal basis $\{ \zeta_1^x, \zeta_2^x\}$ of $\H_x$ such that
\[
U\zeta_1^x \in {\mathbb C} \eta_1^{x} \oplus {\mathbb C} \eta_2^{x+1}
\quad {\rm and} \quad
U\zeta_2^x \in {\mathbb C} \eta_1^{x-1} \oplus {\mathbb C}\eta_2^x .
\]
\end{proposition}
\proof
When ${\rm rank} P_{x} U P_{x+ 1} ={\rm rank} P_{x} U P_{x-1}= 0$, by Lemma  \ref{lem:rank0},
there exist unit vectors  $\zeta_1^x$ and  $\zeta_2^x$ such that
$U\zeta_1^x = \eta_1^x$ and $U\zeta_2^x = \eta_2^x$.
Since $\{\eta_1^x, \eta_2^x\}$ is an orthonormal basis, 
$\{\zeta_1^x, \zeta_2^x\}$ is also an orthonormal basis of $\H_x$.

When ${\rm rank} P_{x} U P_{x+ 1}= {\rm rank} P_{x} U P_{x-1}= 1$,
$U\e_1^x$ and $U\e_2^x$ can be written as
\begin{align*}
U \e_1^x &= a_1 \eta_1^{x-1} + a_2 \eta_1^x + a_3 \eta_2^x + a_4 \eta_2^{x+1} \\
U \e_2^x &= b_1 \eta_1^{x-1} + b_2 \eta_1^x + b_3 \eta_2^x + b_4 \eta_2^{x+1}
\end{align*}
for some $a_i, b_i \in {\mathbb C}$.
Assume that $a_1=b_1=0$. Then, $a_3=b_3=0$ since $U\H_x$ is orthogonal to $U\H_{x-1}$. Hence, $U\H_x = {\mathbb C} \eta_1^x \oplus {\mathbb C}\eta_2^{x+1}$.
However, this contradicts to $U\H_x \perp U\H_{x+1}$ and ${\rm rank}P_x U P_{x+1} =1$.
Therefore, we obtain that $a_1 \neq 0$ or $b_1\neq 0$.
Similarly, we have that $a_4 \neq 0$ or $b_4 \neq 0$.

Let $\zeta_1^x$ be 
the normalization of $b_1\e_1^x - a_1 \e_2^x$,
and let $\zeta_2^x$ be
the normalization of $b_4\e_1^x - a_4 \e_2^x$. 
Then, $\zeta_1^x$ and $\zeta_2^x$ are unit vectors which satisfies
\begin{align*}
U \zeta_1^x &=  \psi_x+ c_1 \eta_2^{x+1}   \\
U \zeta_2^x &=  c_2\eta_1^{x-1} +\phi_x
\end{align*}
for some $\psi_x, \phi_x \in \H_x$ and $c_1, c_2 \in {\mathbb C}$.
Since $U\zeta_1^x$ is orthogonal to $U\H_{x-1}$, $\psi_x$ must be orthogonal to $\eta_2^x$,
and therefore, $\psi_x = d_1 \eta_1^x$ for some $d_1 \in {\mathbb C}$.
Similarly, $\phi_x = d_2 \eta_2^x$ for some $d_2\in {\mathbb C}$.
Consequently, 
\begin{align*}
U \zeta_1^x &=  d_1\eta_1^x +c_1 \eta_2^{x+1} \\
U \zeta_2^x &=  c_2\eta_1^{x-1} + d_2\eta_2^x.
\end{align*}
The vectors on the right hand sides are orthogonal. Hence,
$\{\zeta_1^x, \zeta_2^x\}$ is an orthonormal basis of $\H_x$.

When ${\rm rank} P_x U P_{x+1} =0$ and ${\rm rank} P_x U P_{x-1} =1$,
by Lemma \ref{lem:rank0}, there exists a unit vector $\zeta_1^x$ such that
\[
U\zeta_1^x = \eta_1^x.
\]
Let $\zeta_2^x$ be a unit vector which is orthogonal to $\zeta_1^x$. 
Then, $U\zeta_2^x$ can be written as
\begin{align*}
U \zeta_2^x &= a_1 \eta_1^{x-1} + a_2 \eta_1^x + a_3 \eta_2^x + a_4 \eta_2^{x+1} 
\end{align*}
for some $a_i \in {\mathbb C}$.
Since $U\zeta_1^x \perp U\zeta_2^x$, $a_2=0$.
If $a_4 \neq 0$, $U\H_{x+1}$ must be orthogonal to $\eta_2^{x+1}$.
Hence,
\[
U\left( \bigoplus_{y\neq x} \H_y \right) \perp {\mathbb C} \eta_2^{x+1}.
\]
This means $U\H_x \supset {\mathbb C} \eta_2^{x+1}$, and therefore
$U \zeta_2^x = \alpha \eta_2^{x+1}$.
This contradicts to Assumption {\rm A}.
Thus, we have $a_4=0$.
Consequently, 
\[
U\zeta_2^x =  a_1 \eta_1^{x-1} + a_3 \eta_2^x.
\]

In the case of  ${\rm rank} P_x U P_{x-1} =0$ and ${\rm rank} P_x U P_{x+1} =1$,
we can prove the assertion similarly.
\endproof

Define $\xi_1^x= U\zeta_1^x$ and $\xi_2^x = U\zeta_2^x$.
Then, we obtain the next theorem.

\begin{theorem}\label{thm:str}
For a unitary $U$ on $\H$ with the conditions \eqref{eq:rangecond}, \eqref{eq:rankcond} 
and Assumption {\rm A},
there exist orthonormal bases 
 $\{\eta_1^x, \eta_2^x\}_{x \in {\mathbb Z}}$,
$\{\zeta_1^x, \zeta_2^x\}_{x \in {\mathbb Z}}$ and $\{\xi_1^x, \xi_2^x\}_{x \in {\mathbb Z}}$ of $\H$
such that
$\eta_i^x, \zeta_i^x \in \H_x$,
$\xi_1^x \in   {\mathbb C} \eta_1^x\oplus  {\mathbb C} \eta_2^{x+1}$,
$\xi_2^x \in {\mathbb C} \eta_1^{x-1} \oplus {\mathbb C} \eta_2^{x}$ and 
\[
U = \sum_{x\in {\mathbb Z}} |\xi_1^x\> \< \zeta_1^x| + |\xi_2^x\>\<\zeta_2^x|.
\]
\end{theorem}

\section{Unitary equivalence classes of SSQW}\label{sect3}

In this section, we consider unitary equivalence classes.
\begin{definition}
Two quantum walks $U$ and $U'$ are unitarily equivalent if there exists a
unitary $W = \bigoplus_{x\in {\mathbb Z}} W_x$ on $\H = \bigoplus_{x\in {\mathbb Z}} \H_x$
such that
\[
U' = WUW^*.
\]
\end{definition}

Let $U$ be a unitary on $\H$ with the conditions \eqref{eq:rangecond}, \eqref{eq:rankcond}
and Assumption {\rm A}.
Then, there exist orthonormal bases $\{\eta_1^x, \eta_2^x\}_{x \in {\mathbb Z}}$,
$\{\zeta_1^x, \zeta_2^x\}_{x \in {\mathbb Z}}$ and $\{\xi_1^x, \xi_2^x\}_{x \in {\mathbb Z}}$ of $\H$
with
$\eta_i^x, \zeta_i^x \in \H_x$,
$\xi_1^x \in   {\mathbb C} \eta_1^x\oplus  {\mathbb C} \eta_2^{x+1}$ and
$\xi_2^x \in {\mathbb C} \eta_1^{x-1} \oplus {\mathbb C} \eta_2^{x}$ such that
\[
U = \sum_{x\in {\mathbb Z}} |\xi_1^x\> \< \zeta_1^x| + |\xi_2^x\>\<\zeta_2^x|
\]
by Theorem \ref{thm:str}.
Let $W_1$ be a unitary defined by
\[
W_1 = \bigoplus_{x\in {\mathbb Z}} |\e_1^x\> \<\eta_1^x| + |\e_2^x \> \< \eta_2^x|.
\]
Then, 
\begin{align*}
W_1 U W_1^* 
&= 
\sum_{x\in {\mathbb Z}} |W_1 \xi_1^x\> \< W_1\zeta_1^x| + |W_1\xi_2^x\>\<W_1 \zeta_2^x|.
\end{align*}
Here, $W_1 \xi_1^x \in {\mathbb C} \e_1^{x} \oplus {\mathbb C} \e_2^{x+1}$ and
$W_1 \xi_2^x \in {\mathbb C} \e_1^{x-1} \oplus {\mathbb C} \e_2^{x}$.
Since $\{W_1 \xi_1^{x}, W_1 \xi_2^{x+1}\} $ is an orthonormal basis of 
${\mathbb C} \e_1^{x} \oplus {\mathbb C} \e_2^{x+1}$,
these vectors are parametrized as follows:
\begin{align*}
W_1\xi_1^x &= e^{\im a_{x}} p_{x} \e_1^x + e^{\im b_{x}} q_{x} \e_2^{x+1}  , \\
W_1\xi_2^x &= e^{\im c_{x-1}} q_{x-1} \e_1^{x-1}+ e^{\im d_{x-1}} p_{x-1} \e_2^x  ,\\
W_1\zeta_1^x &= e^{\im \alpha_x} r_x \e_1^x + e^{\im \beta_x} s_x \e_2^x, \\
W_1\zeta_2^x &= e^{\im \gamma_x} s_x \e_1^x + e^{\im \delta_x} r_x \e_2^x,
\end{align*}
where $0\le p_x, r_x \le1$, $q_x =\sqrt{1-p_x^2}$, $s_x =\sqrt{1- r_x^2}$ and
$a_x, b_x, c_x ,d_x, \alpha_x, \beta_x, \gamma_x,$ $\delta_x \in {\mathbb R}$.
If $p_x \neq 0$ and $q_x \neq 0$, $a_x, b_x, c_x$ and $d_x$ should satisfy
\begin{equation}\label{eq:ampli}
a_x - b_x = c_x - d_x +\pi 
\end{equation}
in modulo $2\pi$. When $p_x =0$ (resp. $q_x=0$), $a_x$ and $d_x$ (resp. $b_x$ and $c_x$)
can be arbitrary. Hence, we choose $a_x, b_x, c_x$ and $d_x$ to satisfy \eqref{eq:ampli}.
Similarly, we can assume that $\alpha_x, \beta_x, \gamma_x$ and $\delta_x$ fulfill
\begin{equation}\label{eq:ampli2}
\alpha_x - \beta_x = \gamma_x -\delta_x + \pi
\end{equation}
in modulo $2\pi$.
We will omit the notation ``in modulo $2\pi$'' if there is no confusion.
Then, $W_1UW_1^*$ is written as
\begin{align*}
W_1 U W_1^* 
=& 
\sum_{x\in {\mathbb Z}}
 | e^{\im a_{x}} p_{x} \e_1^{x}+ e^{\im b_{x}} q_{x} \e_2^{x+1}\> 
 \< e^{\im \alpha_x} r_x \e_1^x + e^{\im \beta_x} s_x \e_2^x| \\
& + |e^{\im c_{x-1}} q_{x-1} \e_1^{x-1} + e^{\im d_{x-1}} p_{x-1} \e_2^{x}\>
 \<e^{\im \gamma_x} s_x \e_1^x + e^{\im \delta_x} r_x \e_2^x| \\
 =& 
\sum_{x\in {\mathbb Z}}
| e^{\im (a_{x}-b_x)} p_{x} \e_1^{x}+ q_{x} \e_2^{x+1}\> 
 \< e^{\im (\alpha_x-b_x)} r_x \e_1^x + e^{\im (\beta_x-b_x) } s_x \e_2^x| \\
& + | q_{x-1} \e_1^{x-1} + e^{\im (d_{x-1}-c_{x-1})} p_{x-1} \e_2^{x}\>
 \<e^{\im (\gamma_x-c_{x-1})} s_x \e_1^x + e^{\im (\delta_x-c_{x-1})} r_x \e_2^x| .
\end{align*}
We replace $a_x - b_x$, $\alpha_x -b_x$, $\beta_x -b_x$ 
$d_{x-1} -c_{x-1}$, $\gamma_x -c_{x-1}$ and $\delta_x -c_{x-1}$ by 
$a_x$, $\alpha_x$, $\beta_x$, $d_{x-1}$, $\gamma_x$ and $\delta_x$, respectively. 
Then, 
the conditions \eqref{eq:ampli} and \eqref{eq:ampli2} are rewritten as
\[
a_x = -d_x +\pi \quad {\rm and} \quad \alpha_x - \beta_x = \gamma_x -\delta_x + \pi,
\]
and
\begin{align*}
W_1 U W_1^* 
=& 
\sum_{x\in {\mathbb Z}}
| e^{\im a_{x}} p_{x} \e_1^{x}+ q_{x} \e_2^{x+1}\> 
 \< e^{\im \alpha_x} r_x \e_1^x + e^{\im \beta_x } s_x \e_2^x| \\
& + | q_{x-1} \e_1^{x-1} - e^{-\im a_{x-1}} p_{x-1} \e_2^{x}\>
 \<e^{\im \gamma_x} s_x \e_1^x + e^{\im \delta_x} r_x \e_2^x|.
\end{align*} 

Let $g_x$ and $h_x$ be real numbers defined by
\begin{align}
g_0=0, \quad 
g_x &= -\sum_{k=1}^{x} \gamma_k \quad(x \ge 1), \quad 
g_x = \sum_{k=x+1}^{0} \gamma_k \quad (x \le -1) \nonumber \\
h_0 = \ell , \quad 
h_x &= \ell + \sum_{k=0}^{x-1} \beta_k \quad(x \ge 1), \quad
h_x = \ell - \sum_{k=x}^{-1} \beta_k \quad (x \le -1) \label{ell}
\end{align}
for some $\ell \in {\mathbb R}$.
Note that $g_x -g_{x-1} = -\gamma_x$ and $h_{x+1}-h_x = \beta_x$ for all $x \in{\mathbb Z}$.
Let 
$W_2$ be a unitary operator defined by
\[
W_2 = \bigoplus_{x\in {\mathbb Z}}
\begin{bmatrix}
e^{\im g_x} & 0 \\
0 & e^{\im h_x}
\end{bmatrix}.
\]
By using $\alpha_x + g_x -h_{x+1} = -\delta_x -h_x +g_{x-1} + \pi$ in modulo $2\pi$,
we have 
\begin{align}
&W_2 W_1 U W_1^* W_2^*  \nonumber \\
&=\sum_{x\in {\mathbb Z}}
| e^{\im a_{x}} p_{x} W_2\e_1^{x}+ q_{x} W_2\e_2^{x+1}\> 
 \< e^{\im \alpha_x} r_x W_2\e_1^x + e^{\im \beta_x } s_x W_2\e_2^x| \nonumber \\
& + | q_{x-1} W_2\e_1^{x-1} - e^{-\im a_{x-1}} p_{x-1} W_2\e_2^{x}\>
 \<e^{\im \gamma_x} s_x W_2\e_1^x + e^{\im \delta_x} r_x W_2\e_2^x| \nonumber \\
&= \sum_{x\in {\mathbb Z}}
| e^{\im (a_{x}+g_x - h_{x+1})} p_{x} \e_1^{x}+ q_{x} \e_2^{x+1}\> 
 \< e^{\im (\alpha_x+g_x - h_{x+1})} r_x \e_1^x 
 + e^{\im (\beta_x +h_x -h_{x+1}) } s_x \e_2^x| \nonumber \\
& + | q_{x-1} \e_1^{x-1} - e^{\im (-a_{x-1}+h_x - g_{x-1})} p_{x-1} \e_2^{x}\>
 \<e^{\im (\gamma_x+g_x -g_{x-1})} s_x \e_1^x 
 + e^{\im (\delta_x+ h_x -g_{x-1})} r_x \e_2^x| \nonumber \\
 &=
 \sum_{x\in {\mathbb Z}}
| e^{\im \theta_x} p_{x} \e_1^{x}+ q_{x} \e_2^{x+1}\> 
 \< e^{\im \kappa_x} r_x \e_1^x 
 +s_x \e_2^x| \nonumber \\
& + | q_{x-1} \e_1^{x-1} - e^{-\im\theta_{x-1}} p_{x-1} \e_2^{x}\>
 \<s_x \e_1^x 
 - e^{-\im \kappa_x} r_x \e_2^x|, \label{WUW3}
\end{align}
where 
\begin{equation}\label{thetakappa}
\theta_x = a_x + g_x -h_{x+1} \quad {\rm and}  \quad \kappa_x = \alpha_x+g_x -h_{x+1}.
\end{equation}
Remark that we choose $\ell$ later.

Consequently, we have the next theorem.

\begin{theorem}\label{Upr}
A quantum walk $U$ on $\H$ with conditions \eqref{eq:rangecond}, \eqref{eq:rankcond}
and Assumption {\rm A} is unitarily equivalent to
\begin{align*}
U_{p,r,\theta,\kappa}
=&
 \sum_{x\in {\mathbb Z}}
| e^{\im \theta_x} p_{x} \e_1^{x}+ q_{x} \e_2^{x+1}\> 
 \< e^{\im \kappa_x} r_x \e_1^x 
 +s_x \e_2^x| \\
& + | q_{x-1} \e_1^{x-1} - e^{-\im\theta_{x-1}} p_{x-1} \e_2^{x}\>
 \<s_x \e_1^x 
 - e^{-\im \kappa_x} r_x \e_2^x|
\end{align*}
for some $0 \le p_x, r_x \le 1$ and $\theta_x , \kappa_x \in [0,2\pi)$.
\end{theorem}

When $p_x=0$ (resp. $r_x=0$), we assume $\theta_x=0$ (resp. $\kappa_x =0$).
Remark that unitary equivalence between $U_{p,r,\theta,\kappa}$ and $U_{p',r',\theta',\kappa'}$
does not lead $p=p'$, $r=r'$, $\theta =\theta'$ and $\kappa = \kappa'$, in general.
Hence, to clarify unitary equivalence classes, we need more conditions.

We use the notation $\H_{[y,z]} =\bigoplus_{y \le x \le z} \H_x$ and
$U_{[y,z]} = U |_{\H_{[y,z]}}$ for $y,z \in {\mathbb Z}\cup \{-\infty, \infty\}$.
(If $y = -\infty$, we write $\H_{(-\infty,z]}$, for example.)
When $q_x=0$, $U_{p,r,\theta, \kappa}$ can be decomposed as
\[
U_{p,r,\theta, \kappa} = (U_{p,r,\theta, \kappa})_{(-\infty, x]} 
\oplus (U_{p,r,\theta, \kappa})_{[x+1, \infty)}.
\]
This condition is equivalent to 
${\rm rank} P_{x+1} U_{p,r,\theta, \kappa} P_x = {\rm rank} P_x U_{p,r,\theta, \kappa} P_{x+1} =0$.
Here, we consider the next assumption.

\begin{assumption}
There does not exist $x\in {\mathbb Z}$ such that
\begin{align}
{\rm rank} P_{x} U P_{x-1} = {\rm rank} P_{x-1} U P_x &=0 \quad {\rm and} \nonumber \\
{\rm rank} P_{x+1} U P_x = {\rm rank} P_x U P_{x+1} &=0. \label{B}
\end{align}
\end{assumption}

We use Assumption B for the following reasons. 
When \eqref{B} holds, $U$ can be decomposed as
\[
U = U_{(-\infty, x-1]} \oplus U_x \oplus U_{[x+1, \infty)},
\]
where $U_x = U_{[x,x]}$. Then, $U_x$ is a 2 by 2 unitary matrix.
Since 2 by 2 unitary matrices are well known, 
it is no problem to ignore $U_x$.
Another reason is as follows. Two 2 by 2 unitary matrices are unitarily equivalent 
if and only if their eigenvalues are the same.
Therefore, if $U_{p,r,\theta, \kappa}$ and $U_{p',r',\theta', \kappa'}$
satisfy the conditions $\theta_x = \theta'_x = \kappa_x = \kappa'_x =0$,
$\theta_{x-1}=\theta'_{x-1} =\pi$ and \eqref{B},  the two unitary matrices
\[
(U_{p,r,\theta, \kappa})_x =
\begin{bmatrix}
 r_x & s_x \\
 s_x &  -r_x
\end{bmatrix}
\quad {\rm and} \quad  
(U_{p',r',\theta', \kappa'})_x =
\begin{bmatrix}
 r'_x & s'_x \\
 s'_x &  -r'_x
\end{bmatrix}
\]
are unitarily equivalent for any $r_x$ and $r'_x$.
This makes the calculation complicated, when we clarify unitary equivalence classes.

Now, we consider unitary equivalence between 
$U_{p,r,\theta,\kappa}$ and $U_{p',r',\theta',\kappa'}$.

\begin{lemma}\label{lem:pr}
If two quantum walks
$U_{p,r,\theta,\kappa}$ and $U_{p',r',\theta',\kappa'}$ with Assumption {\rm B}
 are unitarily equivalent,
then $p=p'$, $r=r'$.
\end{lemma}

\proof
Let $U_{p,r,\theta,\kappa}$ and $U_{p',r',\theta',\kappa'}$ be unitarily equivalent.
Then, there exists a unitary $W = \bigoplus_{x\in{\mathbb Z}} W_x$ 
on $\H = \bigoplus_{x\in{\mathbb Z}} \H_x$ such that
\[
U_{p',r',\theta',\kappa'}
=WU_{p,r,\theta,\kappa}W^*.
\]
This means
\begin{align}
&\sum_{x\in {\mathbb Z}}
| e^{\im \theta'_x} p'_{x} \e_1^{x}+ q'_{x} \e_2^{x+1}\> 
 \< e^{\im \kappa'_x} r'_x \e_1^x 
 +s'_x \e_2^x| \nonumber \\
&\quad + | q'_{x-1} \e_1^{x-1} - e^{-\im\theta'_{x-1}} p'_{x-1} \e_2^{x}\>
 \<s'_x \e_1^x 
 - e^{-\im \kappa'_x} r'_x \e_2^x|\nonumber \\
=& \sum_{x\in {\mathbb Z}}
| e^{\im \theta_x}  p_{x} W_x\e_1^{x}+ q_{x} W_x\e_2^{x+1}\> 
 \< e^{\im \kappa_x} r_x W_x\e_1^x 
 +s_x W_x\e_2^x|\nonumber  \\
& \quad + | q_{x-1} W_x\e_1^{x-1} - e^{-\im\theta_{x-1}} p_{x-1} W_x\e_2^{x}\>
 \<s_x W_x\e_1^x 
 - e^{-\im \kappa_x} r_x W_x\e_2^x|. \label{WUW}
 \end{align}
First, we see that $W_x e_i^x \in {\mathbb C}e_i^x$ for all $x\in {\mathbb Z}$ and $i=1,2$.
When $q'_{x} \neq 0$, $P_x U_{p',r',\theta',\kappa'} P_{x+1}
=P_x WU_{p,r,\theta,\kappa}W^* P_{x+1}$ implies
\[
|q'_{x} \e_1^x\>\< s'_{x+1} \e_1^{x+1} - e^{-\im \kappa'_{x+1}} r'_{x+1} \e_2^{x+1}|
=
|q_{x} W_x\e_1^x\>\< s_{x+1} W_x\e_1^{x+1} - e^{-\im \kappa_{x+1}} r_{x+1} W_x\e_2^{x+1}|.
\]
Therefore, we have $q_x \neq 0$ and
\[
{\mathbb C} \e_1^x=
{\rm Ran} P_x U_{p',r',\theta',\kappa'} P_{x+1}
={\rm Ran}P_x WU_{p,r,\theta,\kappa}W^* P_{x+1}
= {\mathbb C} W_x \e_1^x.
\]
Since $W_x$ is unitary, $W_x \e_2^x \in {\mathbb C} \e_2^x$ also holds.
Similarly, when $q'_{x-1}\neq 0$, 
\[
{\rm Ran} P_x U_{p',r',\theta',\kappa'} P_{x-1}
={\rm Ran} P_x WU_{p,r,\theta,\kappa}W^* P_{x-1}
\]
implies
$W_x \e^x_2 \in {\mathbb C} \e^x_2$, 
and hence $W_x \e^x_1 \in {\mathbb C} \e^x_1$.
By Assumption B, at least one of $q'_{x-1}\neq 0$ and $q'_{x}\neq 0$ holds.
Thus, we conclude that 
$W_x e_i^x \in {\mathbb C}e_i^x$ for all $x\in {\mathbb Z}$ and $i=1,2$.
Therefore, $W$ can be written as
\[
W = \bigoplus_{x\in {\mathbb Z}}
\begin{bmatrix}
e^{\im g_x} & 0 \\
0 & e^{\im h_x}
\end{bmatrix}
\]
for some $g_x, h_x \in {\mathbb R}$, and \eqref{WUW} is 
\begin{align}
&\sum_{x\in {\mathbb Z}}
| e^{\im \theta'_x} p'_{x} \e_1^{x}+ q'_{x} \e_2^{x+1}\> 
 \< e^{\im \kappa'_x} r'_x \e_1^x 
 +s'_x \e_2^x| \nonumber \\
&\quad + | q'_{x-1} \e_1^{x-1} - e^{-\im\theta'_{x-1}} p'_{x-1} \e_2^{x}\>
 \<s'_x \e_1^x 
 - e^{-\im \kappa'_x} r'_x \e_2^x|\nonumber \\
=& \sum_{x\in {\mathbb Z}}
| e^{\im \theta_x}  p_{x}  \e_1^{x}+ e^{\im( h_{x+1}-g_x)} q_{x}\e_2^{x+1}\> 
 \< e^{\im \kappa_x} r_x \e_1^x 
 +e^{\im( h_{x}-g_x)} s_x \e_2^x|\nonumber  \\
& \quad + | e^{\im( g_{x-1}-h_{x})} q_{x-1} \e_1^{x-1} - e^{-\im\theta_{x-1}} p_{x-1} \e_2^{x}\>
 \<e^{\im( g_x - h_{x})} s_x \e_1^x 
 - e^{-\im \kappa_x} r_x \e_2^x|. \label{WUW2}
 \end{align}

Next, we show that $p = p'$ and $r=r'$.
Comparing the absolute values of coefficients of $|\e_{1}^x\>\< \e_1^x|$,
$|\e_{1}^x\>\< \e_2^x|$ and $|\e_{2}^{x+1}\>\< \e_1^x|$,
we have
\[
p_xr_x = p'_xr'_x, \quad p_x s_x = p'_x s'_x , \quad {\rm and} \quad q_x r_x = q'_x r'_x.
\]
By straightforward calculation, we obtain $p=p'$ and $r=r'$.
\endproof

As already mentioned,
when $q_x=0$, $U_{p,r,\theta, \kappa}$ can be decomposed as
\[
U_{p,r,\theta, \kappa} = (U_{p,r,\theta, \kappa})_{(-\infty, x]} 
\oplus (U_{p,r,\theta, \kappa})_{[x+1, \infty)}.
\]
Hence, without loss of generality, it is sufficient to consider 
unitary equivalence classes of the following four types of quantum walks:

(1) $U_{p,r,\theta, \kappa}$ with $q_x \neq 0$ for all $x\in{\mathbb Z}$.

(2) $(U_{p,r,\theta, \kappa})_{[1, \infty)}$ with 
$q_0=0$ and 
 $q_x \neq 0$ for all $x\ge 1$.

(3) $(U_{p,r,\theta, \kappa})_{(-\infty, 0]}$ with 
$q_0=0$ and 
$q_x \neq 0$ for all
$x\le -1$.

(4) $(U_{p,r,\theta, \kappa})_{[1, y]}$ for $y\ge 2$ 
with 
$q_0=0$,
 $q_y = 0$ and $q_x \neq 0$ for all $1\le x\le y-1$.

\bigskip
\noindent
{\bf Case (1).}
First, we define $\ell \in {\mathbb R}$ as follows:

(i) When $r_x\neq 0$ for some $x \ge 0$, 
we define $w = {\rm min} \{ x\ge 0 \colon r_x \neq 0\}$ and 
$\ell = \alpha_w + g_w - \sum_{k=0}^w \beta_k$.
This implies $\kappa_w =0$.

(ii) When $r_x =0$ for all $x \ge 0$ and $r_x \neq 0$ for some $x \le -1$,
we define $ w = {\rm max} \{ x\le -1 \colon r_x \neq 0\}$ and
$\ell = \alpha_w +g_w + \sum_{k=w+1}^{-1} \beta_k$.
This implies $\kappa_w =0$.

(iii) When $r_x=0$ for all $x\in{\mathbb Z}$ and $p_x \neq 0$ for some $x \ge 0$, 
we define $w = {\rm min} \{ x\ge 0 \colon p_x \neq 0\}$ and 
$\ell = a_w +g_w - \sum_{k=0}^w \beta_k$. 
This implies $\theta_w =0$.

(iv) When $r_x=0$ for all $x\in{\mathbb Z}$, $p_x =0$ for all $x \ge 0$ and
$p_x \neq 0$ for some $x \le -1$, 
we define $ w = {\rm max} \{ x\le -1 \colon p_x \neq 0\}$ and
$\ell = a_w +g_w + \sum_{k=w+1}^{-1} \beta_k$.
This implies $\theta_w =0$.

(v) When $r_x= p_x=0$ for all $x \in{\mathbb Z}$, we define $w=0$ and $\ell = 0$.

\begin{theorem}
Two quantum walks 
$U_{p,r,\theta,\kappa}$ and $U_{p',r',\theta',\kappa'}$ in Case {\rm (1)}
 are unitarily equivalent if and only if 
$p=p'$, $r=r'$, $\theta=\theta'$ and $\kappa = \kappa'$.
\end{theorem}

\proof
If $p=p'$, $r=r'$, $\theta=\theta'$ and $\kappa = \kappa'$,
then it is obvious that $U_{p,r,\theta,\kappa}$ and $U_{p',r',\theta',\kappa'}$
 are unitarily equivalent.
 
If $U_{p,r,\theta,\kappa}$ and $U_{p',r',\theta',\kappa'}$
are unitarily equivalent, $p=p'$ and $r=r'$ by Lemma \ref{lem:pr}.

Next, we show that $\kappa = \kappa'$. 
When $r_x \neq 0$ for $x \in {\mathbb Z}$, 
comparing coefficients of $|\e_2^{x+1}\>\<\e_1^x|$ and 
$|\e_1^{x-1}\>\<\e_2^x|$ in \eqref{WUW2}, we have
\begin{equation}\label{kappa1}
\kappa'_x = \kappa_x +g_x -h_{x+1}  \quad {\rm and} \quad 
\kappa'_x = \kappa_x +g_{x-1} -h_x
\end{equation}
in modulo $2\pi$. Similarly,  when $s_x \neq 0$, 
comparing coefficients of $|\e_2^{x+1}\>\<\e_2^x|$ and 
$|\e_1^{x-1}\>\<\e_1^x|$ in \eqref{WUW2},
we have 
\begin{equation}\label{eq:gh}
g_{x-1} = g_x \quad {\rm and} \quad  h_x=h_{x+1}.
\end{equation}
In both cases,
\[
g_x -h_{x+1} = g_{x-1} -h_x
\]
holds. 

If $r_x=0$ for all $x \in {\mathbb Z}$, we conclude $\kappa = \kappa' = 0$ by definition.
Hence, we assume that there exists $w \in {\mathbb Z}$ such that $r_w \neq 0$.
Since $\kappa_w = \kappa'_w=0$ by (i) and (ii), 
we get $g_w -h_{w+1} =0$ by \eqref{kappa1}, and therefore,
$g_x - h_{x+1} =0$ for all $x \in {\mathbb Z}$.
When $r_x=0$, we have $\kappa_x = \kappa'_x =0$.
When $r_x \neq 0$, we have $\kappa_x = \kappa'_x$ 
by $g_x-h_{x+1}=0$ and \eqref{kappa1}.
Consequently, we obtain $\kappa = \kappa'$.
Moreover, we get the following:

(a) When $r_x=0$ for all $x \in {\mathbb Z}$, 
$g_x = g_{x+1}$ and $h_x = h_{x+1}$ for all $x \in {\mathbb Z}$.

(b) When $r_x \neq 0$ for some $x \in {\mathbb Z}$, $g_{x} = h_{x+1}$ for all $x \in {\mathbb Z}$.

Finally, we see that $\theta = \theta'$.
If $p_x=0$ for all $x \in {\mathbb Z}$, we conclude $\theta = \theta' = 0$ by definition.
Hence, we assume that there exists $w \in {\mathbb Z}$ such that $p_w \neq 0$.
When $r_x=0$ for all $x \in {\mathbb Z}$, 
$g_x = g_{x+1}$ and $h_x = h_{x+1}$ for all $x \in {\mathbb Z}$ by (a).
Moreover, $\theta_w = \theta'_w = 0$ by (iii) and (iv).
Comparing coefficients of 
$|\e_1^w\>\<\e_2^w|$ in \eqref{WUW2},
$p_w \neq 0$ and $\theta_w = \theta'_{w}=0$ imply 
\[
g_w - h_w =0.
\]
Therefore, $g_x - h_x =0$ for all $x \in {\mathbb Z}$.
When $p_x =0$, we obtain $\theta_x = \theta'_x =0$ by definition.
When $p_x \neq 0$, comparing coefficients of 
$|\e_1^x\>\<\e_2^x|$ in \eqref{WUW2},
we conclude $\theta = \theta'$.

We assume that there exists $y \in {\mathbb Z}$ such that $r_y \neq 0$.
When $p_x =0$, we obtain $\theta_x = \theta'_x =0$ by definition.
Thus, we assume $p_x \neq 0$.
If $r_x\neq 0$, comparing a coefficient of
$|\e_1^x\>\<\e_1^x|$ in \eqref{WUW2},
we get $\theta_x = \theta'_x$.
If $r_x=0$, comparing coefficients of
$|\e_1^x\>\<\e_2^{x}|$ 
 in \eqref{WUW2},
we have
\[
\theta'_{x} = \theta_{x} +g_x -h_x.
\]
Then, by \eqref{eq:gh} and (b), $\theta'_x = \theta_x$.
Consequently, we conclude $\theta = \theta'$.
\endproof

This thoerem says that unitary equivalence classes of quantum walks
with conditions \eqref{eq:rangecond} and
\[
{\rm rank} P_{\pm x} U P_{x} = 1
\]
for all $x \in{\mathbb Z}$ are parametrized by $\{U_{p,r,\theta, \kappa}\}$
with settings (i)-(v).

\bigskip
\noindent
{\bf Case (2).}
When $q_0=0$, $U_{p,r,\theta,\kappa}$ can be decomposed as
\[
U_{p,r,\theta, \kappa} = (U_{p,r,\theta, \kappa})_{(-\infty, 0]} 
\oplus (U_{p,r,\theta, \kappa})_{[1, \infty)}.
\]
In this case, a unitary $W=\bigoplus_{x\in {\mathbb Z}} W_x$ can be also decomposed, and
\begin{align*}
W U_{p,r,\theta, \kappa} W^*
=& W_{(-\infty, 0]} (U_{p,r,\theta, \kappa})_{(-\infty, 0]} W_{(-\infty, 0]}^* \\
&\oplus W_{[1, \infty)} (U_{p,r,\theta, \kappa})_{[1, \infty)} 
W_{[1, \infty)}^*.
\end{align*}
Hence, we restrict our attention to $\H_{[1,\infty)}$.
For convenience, we omit $[1,\infty)$ in this subsection if there is no confusion.

First, we redefine $g_x$ and $h_x$ in \eqref{ell} as
\begin{align*}
g_x &= -\sum_{k=1}^{x} \gamma_k \quad(x \ge 1), \\\
h_1=a_0,  \quad 
h_x &= a_0 + \sum_{k=1}^{x-1} \beta_k \quad(x \ge 2).
\end{align*}
Then, $h_{x+1} -h_x =\beta_x$ and $g_x - g_{x-1} = -\gamma_x$ for $x \ge 1$ with $g_0=0$.
Therefore, $\alpha_x + g_x -h_{x+1} = -\delta_x -h_x +g_{x-1} +\pi$ for $x \ge 1$, and
\begin{align*}
&W_2 W_1 UW_1^* W_2^* \\
&= \sum_{x\ge 1}
| e^{\im (a_{x}+g_x - h_{x+1})} p_{x} \e_1^{x}+ q_{x} \e_2^{x+1}\> 
 \< e^{\im (\alpha_x+g_x - h_{x+1})} r_x \e_1^x 
 + e^{\im (\beta_x +h_x -h_{x+1}) } s_x \e_2^x| \nonumber \\
& + | q_{x-1} \e_1^{x-1} - e^{\im (-a_{x-1}+h_x - g_{x-1})} p_{x-1} \e_2^{x}\>
 \<e^{\im (\gamma_x+g_x -g_{x-1})} s_x \e_1^x 
 + e^{\im (\delta_x+ h_x -g_{x-1})} r_x \e_2^x| \nonumber \\
 &=
 \sum_{x\ge 1}
| e^{\im \theta_x} p_{x} \e_1^{x}+ q_{x} \e_2^{x+1}\> 
 \< e^{\im \kappa_x} r_x \e_1^x 
 +s_x \e_2^x| \nonumber \\
& + | q_{x-1} \e_1^{x-1} - e^{-\im\theta_{x-1}} p_{x-1} \e_2^{x}\>
 \<s_x \e_1^x 
 - e^{-\im \kappa_x} r_x \e_2^x|\\
 &= U_{p,r,\theta,\kappa},
\end{align*}
where $q_0 =\theta_0= 0$, $\theta_x = a_x + g_x -h_{x+1}$ $(x \ge 1)$ and 
$\kappa_x = \alpha_x + g_x -h_{x+1}$ $(x \ge 1)$.

Now, we show the next theorem.

\begin{theorem}
Two quantum walks 
$U_{p,r,\theta,\kappa}$ and $U_{p',r',\theta',\kappa'}$ in Case {\rm (2)}
 are unitarily equivalent if and only if 
$p=p'$, $r=r'$, $\theta=\theta'$ and $\kappa = \kappa'$.
\end{theorem}

\proof
If $p=p'$, $r=r'$, $\theta=\theta'$ and $\kappa = \kappa'$,
it is obvious that $U_{p,r,\theta,\kappa}$ and $U_{p',r',\theta',\kappa'}$ are unitarily equivalent.

If $U_{p,r,\theta,\kappa}$ and $U_{p',r',\theta',\kappa'}$
 are unitarily equivalent, $p=p'$ and $r=r'$ by Lemma \ref{lem:pr}.
 
Next, we show that $\kappa = \kappa'$.
When $r_x \neq 0$ for $x \ge 1$, 
comparing coefficients of $|\e_2^{x+1}\>\<\e_1^x|$ and 
$|\e_1^{x-1}\>\<\e_2^x|$ in \eqref{WUW2}, we have
\begin{equation}\label{kappa2}
\kappa'_x = \kappa_x +g_x -h_{x+1} \quad (x \ge 1)  \quad {\rm and} \quad 
\kappa'_x = \kappa_x +g_{x-1} -h_x \quad (x \ge 2)
\end{equation}
in modulo $2\pi$. Similarly,  when $s_x \neq 0$, we have 
\begin{equation}\label{eq:gh2}
g_{x-1} = g_x \quad (x \ge 2) \quad {\rm and} \quad  h_x=h_{x+1} \quad (x \ge 1).
\end{equation}
In both cases,
\[
g_x -h_{x+1} = g_{x-1} -h_x \quad  (x \ge 2)
\]
holds. 

If $r_1 \neq 0$, comparing coefficients of $|\e_2^1\>\<\e_2^1|$ in \eqref{WUW2},
we get $\kappa_1 = \kappa'_1$. By \eqref{kappa2} with $x=1$, $g_1 -h_2=0$.
Therefore, $g_{x} - h_{x+1} =0$ for all $x \ge 1$.
Similarly, if $s_1 \neq 0$, we get $g_1 =h_1$ and $h_1=h_2$, and thus,
$g_{x} - h_{x+1} =0$ for all $x \ge 1$.

When $r_x =0$, we have $\kappa_x = \kappa'_x=0$.
When $r_x \neq 0$, by $g_x - h_{x+1} =0$ and \eqref{kappa2}, we obtain $\kappa_x = \kappa'_x$.
Hence, we conclude $\kappa = \kappa'$.

Finally, we see $\theta = \theta'$.
If $p_x =0$, then $\theta_x = \theta'_x =0$ by definition.
If $p_x\neq $ and $r_x \neq 0$, comparing coefficients in \eqref{WUW2},
we have $\theta'_x - \kappa'_x = \theta_x - \kappa _x$, and therefore
$\theta'_x = \theta_x$.
If $p_x \neq 0$ and $s_x \neq 0$, comparing coefficients in \eqref{WUW2},
we have $\theta'_x =\theta_x +g_x -h_x$. 
Moreover, by $g_x-h_{x+1}=0$ and \eqref{eq:gh2}, we obtain $g_x - h_x =0$.
Hence, $\theta'_x = \theta_x$.
This completes the assertion.
\endproof

\bigskip
\noindent
{\bf Case (3).}
This case is similar to Case (2). 
Hence, we present the results and omit the details.

A quantum walk $U$ in Case (3) is unitarily equivalent to
\begin{align*}
& \sum_{x\le 0}
| e^{\im \theta_x} p_{x} \e_1^{x}+ q_{x} \e_2^{x+1}\> 
 \< e^{\im \kappa_x} r_x \e_1^x 
 +s_x \e_2^x| \nonumber \\
& + | q_{x-1} \e_1^{x-1} - e^{-\im\theta_{x-1}} p_{x-1} \e_2^{x}\>
 \<s_x \e_1^x 
 - e^{-\im \kappa_x} r_x \e_2^x|\\
 &= U_{p,r,\theta,\kappa},
\end{align*}
where $q_0 =\theta_0= 0$, and the next theorem holds.

\begin{theorem}
Two quantum walks 
$U_{p,r,\theta,\kappa}$ and $U_{p',r',\theta',\kappa'}$ in Case {\rm (3)}
 are unitarily equivalent if and only if 
$p=p'$, $r=r'$, $\theta=\theta'$ and $\kappa = \kappa'$.
\end{theorem}

\bigskip
\noindent
{\bf Case (4).}
This case is also similar to Case (2). 
Hence, we present the results and omit the details.

A quantum walk $U$ in Case (4) is unitarily equivalent to
\begin{align*}
& \sum_{x=1}^y
| e^{\im \theta_x} p_{x} \e_1^{x}+ q_{x} \e_2^{x+1}\> 
 \< e^{\im \kappa_x} r_x \e_1^x 
 +s_x \e_2^x| \nonumber \\
& + | q_{x-1} \e_1^{x-1} - e^{-\im\theta_{x-1}} p_{x-1} \e_2^{x}\>
 \<s_x \e_1^x 
 - e^{-\im \kappa_x} r_x \e_2^x|\\
 &= U_{p,r,\theta,\kappa},
\end{align*}
where $q_0 =q_y= \theta_0= 0$, and the next theorem holds.

\begin{theorem}
Two quantum walks 
$U_{p,r,\theta,\kappa}$ and $U_{p',r',\theta',\kappa'}$ in Case {\rm (4)}
 are unitarily equivalent if and only if 
$p=p'$, $r=r'$, $\theta=\theta'$ and $\kappa = \kappa'$.
\end{theorem}


\section{Chiral symmetry and Suzuki's SSQW}\label{sect4}

\begin{definition}
Let $U$ be a unitary operator of $\H$.
Then, we say that $U$ has chiral symmetry if there exists 
a self-adjoint unitary operator $\Gamma$ on $\H$ such that
\[
\Gamma U \Gamma = U^*.
\]
\end{definition}

The next lemma is known \cite{S2}.

\begin{lemma}
$U$ has chiral symmetry if and only if there exist self-adjoint unitary operators
$\Gamma$ and $C$ on $\H$ such that
\[
U = \Gamma C.
\]
\end{lemma}

Susuki's SSQW is defined as 
\[
U = SC =  \begin{bmatrix} p & qL \\ L^*q^* & -L^*pL \end{bmatrix}
\begin{bmatrix} a & b^* \\ b & - a \end{bmatrix}
\]
for $a_x, p_x \in {\mathbb R}$ and $q_x , b_x \in {\mathbb C}$ 
with $p_x^2 + |q_x|^2 =1$ and $a_x^2 + |b_x|^2 =1$.
By this lemma, it is obvious that $U$ has chiral symmetry.

Here, we consider conditions with which $U_{p,r,\theta, \kappa}$ has chiral symmetry

\begin{proposition}
A unitary $U_{p,r,\theta,\kappa}$ has chiral symmetry, if $\theta = \kappa =0$.
\end{proposition}
\proof
$U_{p,r,\theta,\kappa}$ is expressed as
\begin{align*}
U_{p,r,\theta,\kappa}
=&
 \sum_{x\in {\mathbb Z}}
| e^{\im \theta_x} p_{x} \e_1^{x}+ q_{x} \e_2^{x+1}\> 
 \< e^{\im \kappa_x} r_x \e_1^x 
 +s_x \e_2^x| \\
& + | q_{x-1} \e_1^{x-1} - e^{-\im\theta_{x-1}} p_{x-1} \e_2^{x}\>
 \<s_x \e_1^x 
 - e^{-\im \kappa_x} r_x \e_2^x| \\
&= 
\begin{bmatrix} e^{\im \theta} p & qL \\ L^*q & -L^*e^{\im \theta^*} p L \end{bmatrix}
\begin{bmatrix} e^{\im \kappa^*} r & s \\ s & -e^{\im \kappa} r  \end{bmatrix}\\
&=: SC.
\end{align*}
Therefore, $S$ and $C$ are self-adjoint unitary if $\theta = \kappa =0$.
\endproof

We note that the converse is an open question.

Finally, we consider unitary equivalence classes of Suzuki's SSQW.
When $\theta = \kappa =0$, we write $U_{p,r,\theta,\kappa} = U_{p,r}$.

\begin{corollary}
A Suzuki's SSQW
\[
U =\begin{bmatrix} p & qL \\ L^*q^* & -L^*pL \end{bmatrix}
\begin{bmatrix} a & b^* \\ b & - a \end{bmatrix}
\]
for $a_x, p_x \in {\mathbb R}$ and $q_x , b_x \in {\mathbb C}$ 
with $p_x^2 + |q_x|^2 =1$ and $a_x^2 + |b_x|^2 =1$ is unitarily equivalent to
$U_{p,a}$.
\end{corollary}
\proof
Let $q_x = e^{\im \mu_x} |q_x|$ and $b_x = e^{\im \nu_x}|b_x|$.
Then, $U$ is written as
\begin{align*}
U=&
 \sum_{x\in {\mathbb Z}}
| p_{x} \e_1^{x}+ e^{-\im \mu_x}|q_{x}| \e_2^{x+1}\> 
 \< a_x \e_1^x 
 +e^{\im \nu_x}|b_x| \e_2^x| \\
& + | e^{\im \mu_{x-1}}|q_{x-1}| \e_1^{x-1} -p_{x-1} \e_2^{x}\>
 \<e^{-\im \nu_x}|b_x|\e_1^x 
-a_x \e_2^x| \\
=&
\sum_{x\in {\mathbb Z}}
| e^{\im \mu_x}p_{x} \e_1^{x}+ |q_{x}| \e_2^{x+1}\> 
 \< e^{\im \mu_x}a_x \e_1^x 
 +e^{\im (\nu_x+\mu_x)}|b_x| \e_2^x| \\
& + | |q_{x-1}| \e_1^{x-1} -e^{-\im \mu_{x-1}}p_{x-1} \e_2^{x}\>
 \<e^{\im (-\nu_x-\mu_{x-1})}|b_x|\e_1^x 
-e^{-\im \mu_{x-1}}a_x \e_2^x|.
\end{align*}
Therefore, by \eqref{ell} and \eqref{thetakappa} with $\ell = -\nu_0$, for $x \ge 1$,
\begin{align*}
\theta_x 
&= \mu_x - \sum_{k=1}^x (-\nu_k -\mu_{k-1}) - \ell - \sum_{k=0}^{x} (\nu_k + \mu_k) =0
\quad (x\ge 1)\\
\theta_0 &= \mu_0 -\ell -(\nu_0 + \mu_0) = 0\\
\theta_{-1} &= \mu_{-1} + ( -\nu_0-\mu_{-1}) -\ell =0 \\
\theta_x &= \mu_x +\sum_{k=x+1}^0 (-\nu_k -\mu_{k-1}) -\ell 
+ \sum_{k=x+1}^{-1} (\nu_k +\mu_k) =0 \quad (x\le -2).
\end{align*}
Hence, $\theta =0$. Similarly, $\kappa=0$.
Consequently, we conclude that $U$ is unitarily equivalent to $U_{p,a}$ by 
Theorem \ref{Upr}.
\endproof

This corollary says that we can assume that $q_x , b_x \in {\mathbb R}$ 
in the definition of Suzuki's SSQW.

\begin{corollary}
Two quantum walks $U_{p,a}$ and $U_{p', a'}$ are
unitarily equivalent if and only if $p=p'$ and $a=a'$.
\end{corollary}
\proof
Most parts of the proof are already shown in the theorems in Sect. \ref{sect3}.
The remaining part is the case that 
$U_{p,a}$ and $U_{p', a'}$ do not satisfy Assumption B.

Assume that $U_{p,a}$ and $U_{p', a'}$ do not satisfy Assumption B at $x \in {\mathbb Z}$.
Then, since $p_x= p'_x=1$, $(U_{p,a})_x$ is
\[
(U_{p,a})_x = \begin{bmatrix} a_x & b_x \\ -b_x & a_x \end{bmatrix},
\]
where $b_x = \sqrt{1-a_x^2}$.
Therefore, eigenvalues of $(U_{p,a})_x$ and $(U_{p',a'})_x$ are the same
if and only if $a_x =a'_x$.
This completes the proof.
\endproof

\section{Conclusion}
We study a generalization of SSQW.
In Sect. \ref{sect2}, we show that a quantum walk $U$ 
with the conditions \eqref{eq:rangecond}, \eqref{eq:rankcond} 
and Assumption {\rm A} can be expressed as
\[
U = \sum_{x\in {\mathbb Z}} |\xi_1^x\> \< \zeta_1^x| + |\xi_2^x\>\<\zeta_2^x|,
\]
where 
 $\{\eta_1^x, \eta_2^x\}_{x \in {\mathbb Z}}$,
$\{\zeta_1^x, \zeta_2^x\}_{x \in {\mathbb Z}}$ and 
$\{\xi_1^x, \xi_2^x\}_{x \in {\mathbb Z}}$ are orthonormal bases of $\H$
with
$\eta_i^x, \zeta_i^x \in \H_x$,
$\xi_1^x \in   {\mathbb C} \eta_1^x\oplus  {\mathbb C} \eta_2^{x+1}$ and
$\xi_2^x \in {\mathbb C} \eta_1^{x-1} \oplus {\mathbb C} \eta_2^{x}$.

In Sect. \ref{sect3}, we consider unitary equivalence of such quantum walks.
The above $U$ is unitarily equivalent to 
\begin{align*}
U_{p,r,\theta,\kappa}
=&
 \sum_{x\in {\mathbb Z}}
| e^{\im \theta_x} p_{x} \e_1^{x}+ q_{x} \e_2^{x+1}\> 
 \< e^{\im \kappa_x} r_x \e_1^x 
 +s_x \e_2^x| \\
& + | q_{x-1} \e_1^{x-1} - e^{-\im\theta_{x-1}} p_{x-1} \e_2^{x}\>
 \<s_x \e_1^x 
 - e^{-\im \kappa_x} r_x \e_2^x|
\end{align*}
for some $0 \le p_x, r_x \le 1$ and $\theta_x , \kappa_x \in [0,2\pi)$.
(When $p_x=0$ (resp. $r_x=0$), we assume that $\theta_x=0$ (resp. $\kappa_x =0$).)
The explicit form of the unitary equivalence classes of such quantum walks is 
a little complicated, but roughly speaking, we need four real parameters 
(or two complex parameters) for each vertex $x \in {\mathbb Z}$.
In addition, we can erase one real parameter $\theta_w$ or $\kappa_w$.

In Sect. 4, we clarify the unitary equivalence classes of Suzuki's SSQW.
We also consider chiral symmetry, 
and we show that a quantum walk $U_{p,r,\theta,\kappa}$ has
chiral symmetry if it is a Suzuki's SSQW.
The converse is an open question.



\end{document}